\documentclass[aps,prb,twocolumn,superscriptaddress]{revtex4}
\usepackage{amsmath}
\usepackage{amssymb}
\usepackage{graphicx}

\begin{document}

\title{Quantum size effects in Pb layers with absorbed Kondo adatoms:
Determination of the exchange coupling constant}

\author{U. Schwingenschl\"ogl}
\affiliation{KAUST, PCSE Division, P.O. Box 55455, Jeddah 21534, Saudi Arabia}
\affiliation{Institut f\"ur Physik, Universit\"at Augsburg, D-86135, Augsburg, Germany}
\author{I. A. Shelykh}
\affiliation{Science Institute, University of Iceland, Dunhagi 3, IS-107, Reykjavik, Iceland}

\date{\today}

\begin{abstract}
We consider the magnetic interaction of manganese phtalocyanine (MnPc) absorbed on
Pb layers which were grown on a Si substrate. We perform an ab-initio calculation of the
density of states and Kondo temperature as function of the number of Pb monolayers.
Comparison to experimental data [Phys.\ Rev.\ Lett.\ 99, 256601 (2007)] then allows us
to determine the exchange coupling constant $J$ between the spins of the adsorbed
molecules and those of the Pb host. This approach gives rise to a general and reliable
method for obtaining $J$, by combining experimental and numerical results.
\end{abstract}

\maketitle

Strong correlation phenomena built the fundament of a number of
extraordinary effects observed in condensed matter
systems. One of the most drastic manifestations is the Kondo effect
\cite{acH} arising from strong coupling between single localized
spins and the conduction electrons. For an antiferromagnetic-type
interaction below the characteristic Kondo temperature, $T_K$, this
coupling cannot be treated perturbatively. It comes along with the
formation of a narrow peak in the density of states (DOS) of a Kondo
adatom pinned to the Fermi energy of the conduction
electrons, $\varepsilon_F$. Its lineshape can be described by the well-known
Doniach-Sunjic formula \cite{Oliveira,Vivaldo}.

The appearance of the Kondo resonance has various physical
consequences. In the context of the transport properties of bulk
diluted magnetic alloys it manifests as enhancement of the
resistance at low temperatures. In mesoscopic systems consisting of
a single quantum dot coupled to a pair of one-dimensional
leads \cite{dgg98,smc98} it results in a drastic enhancement of the
conductance below $T_K$ \cite{fs99,mk02}.

A system which recently has attracted much attention, both from the
experimental and theoretical side, is composed of a metallic host
surface with a Kondo adatom (adsorbed atom) and a STM tip, non- or
ferromagnetic \cite{acs08}. Contrasting the case of a single quantum
dot coupled to leads, such a system demonstrates a much greater
variety of transport regimes. This arises from the possibility of quantum
interference between distinct transport paths, namely direct
tip-host tunneling and indirect tip-adatom-host tunneling
\cite{Madhavan}. In particular, the latter process gets extremely
efficient for temperatures below $T_{K}$ due to the appearance of a
Kondo resonance in the adatom spectral function. The quantum
interference mechanism leads to an asymmetry of the Kondo
resonance. Instead of a Lorentzian lineshape it acquires a Fano-type lineshape,
which is characteristic of interference between a discret level and a
continuous spectrum. The appearence of this kind of a Fano-Kondo resonance has
been found in various experiments with nonmagnetic tips
\cite{Madhavan,nk02,kn02,pw05} and has been discussed in a number of
theoretical works \cite{as00,mp01,ou00}.

In a recent experimental work the quantum size effect on the Kondo
temperature was investigated by Y.-S.\ Fu et al.\ \cite{FuPRL}. The system
under study consisted of individual manganese phtalocyanine (MnPc)
molecules absorbed on ultrathin Pb films grown on Si(111).
Scanning tunneling spectroscopy shows asymmetric Fano-Kondo peaks in the differential conductance.
One central implication of the work by Y.-S.\ Fu et al.\ \cite{FuPRL} is that the variation of the
film thickness, given by the number of Pb monolayers (MLs), leads to drastic
changes in $T_K$, more specifically to pronounced oscillations with a period
of 2 MLs. The lowest (23 K) and highest (419 K) values occured
at thicknesses of 15 and 17 MLs, respectively. Because the Kondo temperature
scales with the electronic DOS at the Fermi energy,
$\rho={\rm DOS}(\varepsilon_F)$, according to
\begin{equation} \label{Tk}
T_K=A\cdot e^{-1/(J\cdot\rho)},
\end{equation}
where $A$ is an amplitude and $J$ is the exchange coupling constant,
this observation has been attributed to oscillations in the Pb DOS with the
film thickness. Indeed, for every 2 MLs increase of the thickness one
empty quantization band moves down below the Fermi level and gets occupied \cite{Wei}.

In the following we present a state-of-the-art ab-initio investigation
of the electronic structure of Pb films on a Si(111) surface. Determining
$\rho={\rm DOS}(\varepsilon_F)$ as a function of the film thickness and using
Eq.\ \ref{Tk} we can perform a fit to the experimental data given in
Fig.\ 2 of Ref.\ \onlinecite{FuPRL}. Besides the amplitude $A$, the exchange
coupling constant $J$ is the only fitting parameter and therefore can be
obtained without any further assumption.

Our analysis is based on the augmented spherical wave
approach, applying density functional theory within the local density
approximation \cite{aswbook}. This method is particularly suitable for
dealing with unit cells containing many atoms \cite{us1,us2,us3}, which
are needed to describe interfaces, since it utilizes a minimal atomic-like
basis set. For the present system this basis set consists of Si $3s$, $3p$,
$3d$ as well as Pb $6s$, $6p$, $5d$ states, and is complemented by orbitals
of additional augmentation spheres. Brillouin zone integrations are performed
using the linear tetrahedron method, where we checked the convergence of the
calculation with respect the fineness of the k-mesh by means of a growing number
of up to 231 k-points in the irreducible wedge of the supercell Brillouin zone.

Our Si(111)/Pb(111) supercells comprise a Si slab of about 25 \AA\ thickness, an
attached Pb slab ranging from 1 to 18 Pb MLs, and a vacuum slab on top. For generating
the supercells we have started from the hexagonal representation of the fcc
lattice, with the (111) direction along the hexagonal $c_{\rm hex}$ axis. For
the Pb slab we use the experimental lattice constant $a_{\rm Pb}=4.95$ \AA. Moreover,
for attaching the Si slab ($a_{\rm Si}=5.43$ \AA) to the Pb slab we have to artificially
shrink the former and introduce a lattice strain. Although this lattice strain
in the Si slab is substantial, no drawback on the Pb electronic structure
is to be expected since there are no Si states in the vicinity of the Fermi energy.
Our supercells extend 121.25 \AA\ along $c_{\rm hex}$ and have as basal plane a
parallelogram with a 60$^\circ$ interior angle and edges of length $a_{\rm Pb}$.

\begin{figure}[t]
\centering
\includegraphics[width=0.39\textwidth,clip]{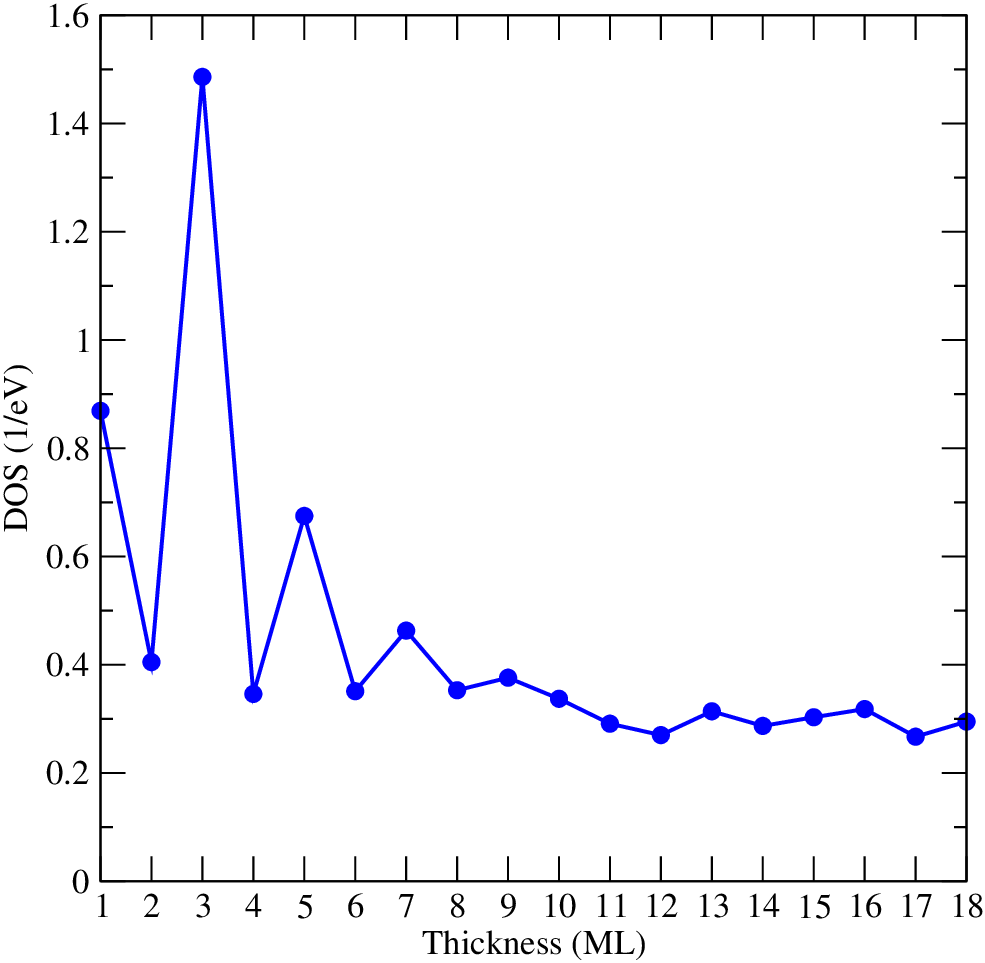}\\\vspace{0.2cm}
\includegraphics[width=0.39\textwidth,clip]{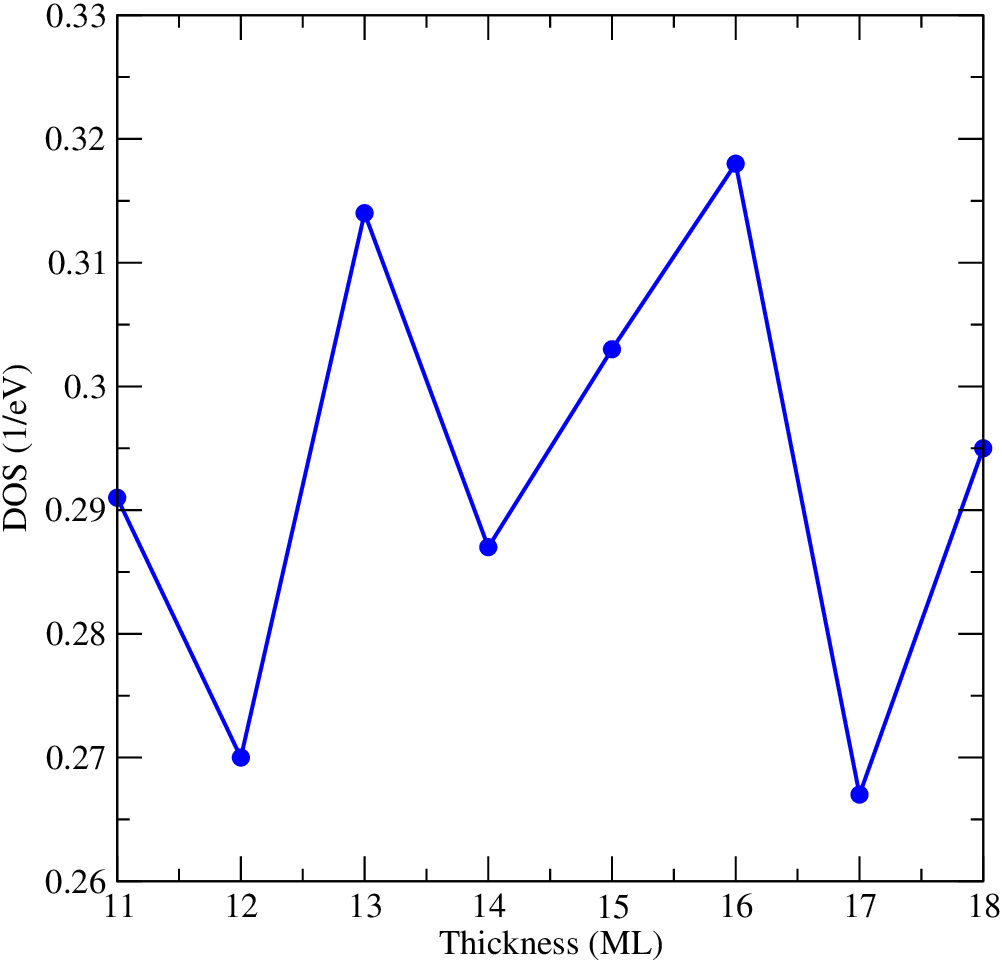}
\caption{Top: Total DOS at $\varepsilon_F$ for a Pb(111) film on Si(111),
depicted as a function of the number of Pb MLs. The values are normalized
by the numbers of Pb atoms in the supercells. Bottom: Zoom for the 11-18 MLs range.}
\label{fig:DOS}
\end{figure}

\begin{figure}[t]
\centering
\includegraphics[width=0.39\textwidth,clip]{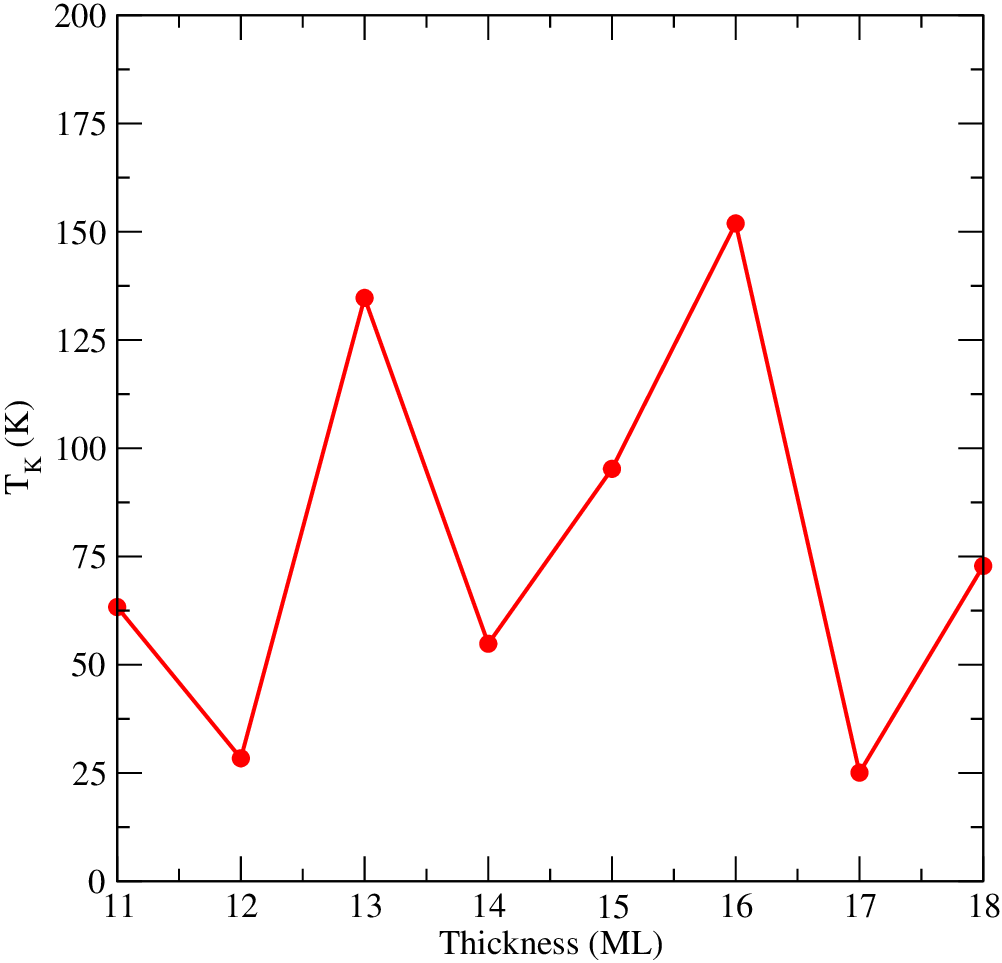}
\caption{Kondo temperature as a function of the thickness of the Pb(111) film,
given by the number of Pb MLs.}
\label{fig:TK}
\end{figure}

We obtain for the DOS at the Fermi energy the results displayed in Fig.\ \ref{fig:DOS}
as a funtion of the Pb film thickness. One clearly identifies the previously discussed
oscillatory behavior, with a period of 2 MLs, up to a thickness of 9 MLs.
Deviations from this regular scheme are found in the intervals 9-12 and 14-16 MLs
where the oszillations are suppressed \cite{envelope}. Comparing Fig.\ \ref{fig:DOS} to
the experimental data \cite{FuPRL} indicates that our curve is shifted by 1 ML to the left hand side,
see particularly the position of the 14-16 MLs anomaly. This difference might either
trace back to the experimental difficulty in counting the MLs or to the effects of
the (planar) MnPc molecule on the local DOS, which is not included in our treatment.
Since there is no ambiguity in the choice of the matching conditions at the Si--Pb
interface, our approach elsewise should not lead to a 1 ML shift.

Using Eq.\ \ref{Tk} we next calculate from our DOS data the dependence of the Kondo
temperature on the film thickness. To that aim, we use the second highest
(150 K at 12 MLs) and smallest (23 K at 15 MLs) of the experimentally observed
$T_K$ values to fit the free parameters $A$ and $J$. The correctness of the
highest measured temperature (419 K at 15 MLs) is quite dubious, see the discussion
of the authors themselves, and thus not taken into consideration. At the
Fermi energy, the ab-initio DOS amounts to 0.318 eV$^{-1}$ for 12 MLs and to
0.267 eV$^{-1}$ for 15 MLs. As a consequence, we obtain from Eq.\ \ref{Tk} for
the exchange coupling constant
\[
J\approx0.32\pm0.04\mbox{ eV}.
\]
As compared to semiconductor bulk systems \cite{prb56}, the value of $J$ is 2--4 times
smaller, which is not surprising because the coupling to a surface in general
is much weaker than the coupling within the bulk.

Due to the logarithmic dependence on $T_K$, the determined value of $J$ is particularly
insensitive to experimental inaccuracies. The outcome of our fitting procedure is displayed
in Fig.\ \ref{fig:TK}. As to be expected, the dependence of $T_K$ on the film thickness
largely resembles the behaviour of $\rho$, see the bottom of Fig.\ \ref{fig:DOS}.
We do not find indications of a dependence of the exchange coupling constant $J$ on the film
thickness between 11 and 18 MLs. Hence, there is a qualitative agreement with the
experimental results, even though clarification of the 1 ML shift and $T_K$ spike at 17 MLs
should be tackled by future experiments.

In summary, we have performed first-principles calculations to establish
the density of states of Pb(111) films grown on a Si(111) substrate. These ab-initio
data have enabled us to fit experimental results from measurements addressing
the dependence of the Kondo temperature on the Pb film thickness. In particular,
we have succeeded in calculating the exchange coupling constant $J$, which is
not accessible to direct experimental determination. For the system under investigation,
a value of about $J=1/3$ eV transpires.

\section*{Acknowledgements}
We thank A.C.\ Seridonio for helpful discussions.


\begin{thebibliography}{99}

\bibitem{acH} A.C.\ Hewson, {\it The Kondo Problem to Heavy Fermions} (Cambridge
University Press, Cambridge, 1993).

\bibitem{Oliveira} H.O.\ Frota and L.N.\ Oliveira, Phys.\ Rev.\ B \textbf{33}, 7871 (1986).

\bibitem{Vivaldo} V.L.\ Campo, Jr.\ and L.N.\ Oliveira, Phys.\ Rev.\ B \textbf{68}, 035337 (2003).

\bibitem{um90} U.\ Meirav, M.A.\ Kastner, and S.J.\ Wind, Phys.\ Rev.\ Lett.\ \textbf{65}, 771 (1990).

\bibitem{dgg98} D.G.\ Gordon, H.\ Shtrikman, D.\ Mahalu, D.A.\ Magder, U.\ Meirav,
M.A.\ Kastner, Nature \textbf{391}, 156 (1998).

\bibitem{smc98} S.M.\ Cronenwett, T.H.\ Oosterkamp, L.P.\ Kouwenhoven, Science \textbf{281}, 540 (1998).

\bibitem{fs99} F.\ Simmel, R.H.\ Blick, J.P.\ Kotthaus, W.\ Wegscheider, M.\ Bichler,
Phys.\ Rev.\ Lett.\ \textbf{83}, 804 (1999).

\bibitem{mk02} M.\ Krawiec and K.I.\ Wysoki\'{n}ski, Phys.\ Rev.\ B \textbf{66}, 165408 (2002).

\bibitem{acs08} A.C.\ Seridonio, F.M.\ Souza, I.A.\ Shelykh, J.Phys.:\ Condens.\ Matter
\textbf{21}, 95003 (2009).

\bibitem{Madhavan} V.\ Madhavan, W.\ Chen, T.\ Jamneala, M.F.\ Crommie, N.S.\ Wingreen,
Science \textbf{280}, 567 (1998).

\bibitem{nk02} N.\ Knorr, M.A.\ Schneider, L.\ Diekh\"oner, P.\ Wahl,
and K.\ Kern, Phys.\ Rev.\ Lett.\ \textbf{88}, 096804 (2002).

\bibitem{kn02} K.\ Nagaoka, T.\ Jamneala, M.\ Grobis, M.F.\ Crommie, Phys.\ Rev.\ Lett.\
\textbf{88}, 077205 (2002).

\bibitem{pw05} P.\ Wahl, L.\ Diekh\"oner, G.\ Wittich, L.\ Vitali, M.A.\ Schneider,
K.\ Kern, Phys.\ Rev.\ Lett.\ \textbf{95}, 166601 (2005).

\bibitem{FuPRL} Y.-S.\ Fu, S.-H.\ Ji, X.\ Chen, X.-C.\ Ma, R.\ Wu, C.-C.\ Wang, W.-H.\
Duan, X.-H.\ Qiu, B.\ Sun, P.\ Zhang, J.-F.\ Jia, Q.-K.\ Xue, Phys.\ Rev.\ Lett.\
\textbf{99}, 256601 (2007).

\bibitem{as00} A.\ Schiller, S.\ Hershfield, Phys.\ Rev.\ B \textbf{61}, 9036 (2000).

\bibitem{mp01} M.\ Plihal, J.W.\ Gadzuk, Phys.\ Rev.\ B \textbf{63}, 085404 (2001).

\bibitem{ou00} O.\ \'Ujs\'aghy, J.\ Kroha, L.\ Szunyogh, A.\ Zawadowski,
Phys.\ Rev.\ Lett.\ \textbf{85}, 2557 (2000).

\bibitem{Wei} C.M.\ Wei and M.Y.\ Chou, Phys.\ Rev.\ B \textbf{66}, 233408 (2002).

\bibitem{aswbook}
V.\ Eyert, {\em The augmented spherical wave method -- a comprehensive treatment},
Lect.\ Notes Phys.\ {\bf 719} (Springer, Heidelberg, 2007).

\bibitem{us1}
U.\ Schwingenschl\"ogl, V.\ Eyert, U.\ Eckern, Europhys.\ Lett.\ {\bf 61}, 361 (2003).

\bibitem{us2}
U.\ Schwingenschl\"ogl and C.\ Schuster, J.\ Phys.:\ Condens.\ Matter \textbf{20}, 382201 (2008);
EPL \textbf{81}, 17007 (2008).

\bibitem{us3}
U.\ Schwingenschl\"ogl, R.\ Fr\'esard, and V.\ Eyert, Surf.\ Sci.\ \textbf{603}, L19 (2009).

\bibitem{envelope}
P.\ Czoschke, H.\ Hong, L.\ Basile, and T.-C.\ Chiang, Phys.\ Rev.\ B {\bf 72}, 075402 (2005).

\bibitem{prb56}
T.\ Mizokawa and A.\ Fujimori, Phys.\ Rev. B {\bf 56}, 6669 (1997).

\end{thebibliography}
\end{document}